\documentclass[hyper]{JHEP3}
\usepackage{amsmath,amssymb,amsfonts,amsxtra,makeidx,graphics,graphicx,amsthm,epsfig,verbatim}
\usepackage{slashed}
\usepackage{bbm}

\newcommand{\be}{\begin{equation}}
\newcommand{\ee}{\end{equation}}
\newcommand{\beq}{\begin{equation}}
\newcommand{\eeq}{\end{equation}}
\newcommand{\ba}{\begin{array}}
\newcommand{\ea}{\end{array}}
\newcommand{\bea}{\begin{eqnarray}}
\newcommand{\eea}{\end{eqnarray}}
\newcommand{\ben}{\begin{enumerate}}
\newcommand{\een}{\end{enumerate}}
\newcommand{\bean}{\begin{eqnarray*}}
\newcommand{\eean}{\end{eqnarray*}}

\newcommand{\Bigsbrk}[1]{\Bigl[#1\Bigr]}
\newcommand{\diag}{\mathop{\mathrm{diag}}}

\newcommand{\nn}{\nonumber}

\newcommand{\tr}{\mathop{\rm Tr}}
\newcommand{\Reals}{\mathbbm{R}}

\newcommand{\trans}{{\scriptscriptstyle\mathrm{T}}}

\newcommand{\BC}{\mathbb{C}}
\newcommand{\BR}{\mathbb{R}}
\newcommand{\BZ}{\mathbb{Z}}
\newcommand{\BN}{\mathbb{N}}

\newcommand{\sD}{\mathcal{D}}

\newcommand{\sV}{\mathcal{V}}

\newcommand{\CC}{{\cal C}}

\newcommand{\BI}{\mathbb{I}}

\newcommand{\ud}{\mathrm{d}}

\newcommand{\Integers}{\mathbbm{Z}}

 \title{M2-branes and AdS/CFT\footnote{Based on lectures delivered by I.R.K. at the Galileo Galilei Institute summer school ``New Perspectives in String Theory.''}}
\author{
\textrm{Igor R.\ Klebanov$^{a,b}$}, \vspace{10mm} \textrm{Giuseppe Torri$^c$}\\
\textit{
$^a$Joseph Henry Laboratories and $^b$Princeton Center for Theoretical Science \\
Princeton University, Princeton, NJ 08544, USA
} \\
\\
\textit{
$^c$Theoretical Physics Group, The Blackett Laboratory \\
Imperial College London, Prince Consort Road\\
London,  SW7 2AZ,  UK
} \\

\texttt{\\ klebanov@princeton.edu, giuseppe.torri08@imperial.ac.uk}

}
\abstract{These notes provide a brief introduction to the ABJM theory, the level $k$ $U(N)\times U(N)$ superconformal Chern-Simons matter theory which has been conjectured to describe $N$ coincident M2-branes. We discuss its dual formulation in terms of M-theory on $AdS_4\times S^7/\Integers_k$ and review some of the evidence in favor of the conjecture. We end with a brief discussion of the important role played by the monopole operators.}

\preprint{PUPT-2313\\
Imperial/TP/09/GT/03}


\begin{document}

\section{Introduction}

 In these notes we review the important progress that has taken place in understanding the world volume theory of coincident supermembranes of M-theory, the M2-branes. These interacting superconformal theories are expected to be the Chern-Simons gauge theories coupled to massless matter \cite{Schwarz:2004yj}. During the past two years, remarkable new theories with extended ${\mathcal N}\geq 6$ superconformal invariance have been constructed \cite{Bagger:2006sk,Gustavsson:2007vu,Aharony:2008ug}. In particular, the ABJM theory \cite{Aharony:2008ug} is a $U(N)\times U(N)$ Chern-Simons gauge theory with integer levels $(k,-k)$; it becomes weakly interacting for $k \gg N$. Considerable amount of evidence has accumulated that this gauge theory describes the low-energy behavior of $N$ M2-branes placed at the ${\mathcal N}=6$ supersymmetric orbifold $\Reals^8/\Integers_k$. Therefore, it is conjectured to be dual, in the sense of the AdS/CFT correspondence \cite{Maldacena:1997re,Gubser:1998bc,Witten:1998qj}, to M-theory on $AdS_4\times S^7/\Integers_k$. This dual description becomes very useful for $N\gg k$.

Before we discuss the role that M2-branes play in the AdS/CFT correspondence, let us review the correspondence for coincident D3-branes, which in flat space realize the well-known $\mathcal{N}=4$ supersymmetric Yang-Mills theory (for reviews see for example \cite{Aharony:1999ti,Klebanov:2000me,D'Hoker:2002aw, Nastase:2007kj, Benna:2008yg}).
The type IIB supergravity background created by a stack of $N$ coincident D3-branes can be written as:
\bea
d s^2 &=& h(r)^{-1/2}(-dt^2 + dx^2_1 + dx^2_2 + dx^2_3) + h(r)^{1/2}(dr^2 + r^2 d\Omega^2_5),\nn \\
h(r) &=& 1 + \frac{L^4}{r^4},\;\;\;\;\;\;L^4 = 4\pi g_s N \alpha'^2,\nn \\
g_s F_5 &=& (1 + *)\; d^4 x \wedge d h^{-1}(r).
\label{n4sym}
\eea
As can be seen from the metric, in the ``near horizon limit,'' which corresponds to taking $r \ll L$, the term $h(r)$ can be approximated by $L^4/r^4$ and, therefore, it is easy to see that the geometry of space-time becomes that of $AdS_5 \times S^5$. Thus, the system of $N$ D3-branes in the near horizon limit is described by a Type IIB string theory in $AdS_5 \times S^5$ space-time.

Now, let us look at this system from the point of view of the world volume theory. Considering only one D3-brane in flat space, one can see that in the low energy limit it is a free gauge theory in $3+1$ dimensions characterized by the lagrangian:
\bea
\mathcal{L} = -\frac{1}{4} F^2_{\mu\nu} - \frac{1}{2} \sum^6_{i=1} (\partial_{\mu} \phi^i)^2 + \ldots,
\eea
where the dots indicate the fermionic terms. This lagrangian corresponds to the $\mathcal{N}=4$ supersymmetric $U(1)$ gauge theory. Since all the fields are uncharged, this is a free theory, which implies that the moduli space\footnote{From the point of view of gauge theory, the moduli space can be thought of as the space of zero-energy solutions of the F-term and D-term equations modded out by the gauge symmetry; from the string theory point of view, the moduli space can be thought of as the space that the branes probe.} is simply $\BR^6$.

This description can be generalized to a system of $N$ parallel D3-branes, where in the low energy limit we have the $\mathcal{N}=4$ supersymmetric Yang-Mills theory with gauge group $U(N)$ and lagrangian:
\bea
\mathcal{L} = \tr\left[-\frac{1}{4} F^2_{\mu\nu} - (D_{\mu} \phi^i)^2 +\frac{g^2}{4} [\phi^i, \phi^j]^2\right] +\ldots,
\eea
where the scalar fields $\phi^i$ transform in the adjoint representation of the gauge group.
The moduli space of this theory consists of diagonal matrices:
\bea
\phi^i = \diag(\phi^i_1, \phi^i_2,\ldots,\phi^i_N),
\eea
with the gauge transformations acting on the fields by a permutation of the eigenvalues. From this, it can be understood that the moduli space of this theory is simply $(\BR^6)^N /S_N$, where $S_N$ is the permutation group of $N$ elements.

There is a $U(1)$ subgroup under which all the fields are neutral; it decouples from the remaining $SU(N)$ gauge group. This $U(1)$ parameterizes the motion of the center of mass of the stack of $N$ D3-branes. Thus, we can conclude that, in the low energy limit, the gauge theory living on a stack of $N$ D3-branes is an $SU(N)$ supersymmetric Yang-Mills theory, which can be studied perturbatively in great detail.
Therefore, a stack of $N$ D3-branes in flat space can be looked at from two points of view: that of string theory which, in the near horizon limit, leads to Type IIB string theory in $AdS_5\times S^5$ with $N$ units of the self-dual 5-form Ramond-Ramond flux, and that of gauge theory that, in the low energy limit, leads to an $\mathcal{N}=4$ super Yang-Mills theory with gauge group $SU(N)$ and adjoint matter. The strong version of the AdS/CFT conjecture claims that these two limits are equivalent. Therefore, the 4-d gauge fields and the 10-d strings provide two different descriptions of the same theory.

 The correspondence may be generalized to the cases where a stack of parallel D3-branes is placed at the tip of a Ricci-flat cone. For this type of background, the metric has the same form as (\ref{n4sym}), but now $d\Omega_5$ refers to the base of the cone, which is an Einstein manifold often denoted as $Y^5$; the limit where $r\rightarrow 0$ gives the geometry of $AdS_5 \times Y^5$.
When the cone is a Calabi-Yau space, the base of the cone is called a Sasaki-Einstein manifold; in these cases the supersymmetry preserved by the theory is one quarter of that of $AdS_5 \times S^5$ that we mentioned above. Hence, these are ${\mathcal N}=1$ superconformal theories.

The  example we will now consider is when the Calabi-Yau cone is the \textit{conifold} (reviews on the conifold theory can be found in \cite{Benna:2008yg, Herzog:2002ih}). This case is of interest here because its gauge theory is related to that appearing in the AdS/CFT duality with M2-branes.
 Let us consider the 10 dimensional space-time $M_4 \times \CC$, where $M_4$ is the usual 4 dimensional Minkowski space-time and $\CC$ is the conifold. Introducing 4 complex variables $u_i$, with $i=1,\ldots,4$, this manifold can be defined as the locus where the following condition is satisfied:
\bea
u^2_1 + u^2_2 + u^2_3 + u^2_4 = 0.
\label{coni}
\eea
Given this condition, it is clear that the conifold has 3 complex dimensions and, therefore, 6 real dimensions.
If we place a D3-brane near the tip of the complex cone defined by (\ref{coni}), we can examine the gauge theory arising on the brane in the low energy limit. This is an $\mathcal{N}=1$ supersymmetric Yang-Mills theory with gauge group $U(1)_1\times U(1)_2$ and four chiral fields, that we shall denote as $z_1, z_2, w_1$ and $w_2$ \cite{Klebanov:1998hh}. They transform under the gauge groups according to Table \ref{t:chargeconi}.\\
\begin{table}[h!]
 \begin{center}  
  \begin{tabular}{|c||c|c|}
  \hline
  \;& $U(1)_1$&$U(1)_2$\\
  \hline\hline  

  $z_1, z_2$&$  1$&$ -1$ \\
  \hline
  $w_1, w_2$&$ -1$&$  1$ \\
  \hline

   \end{tabular}
  \end{center}
\caption{Abelian charges of the chiral fields of the conifold theory under the two gauge groups.}
\label{t:chargeconi}
\end{table}

Interestingly, all the matter fields are neutral under the diagonal gauge field\footnote{We adopt the convention that non hatted quantities refer to gauge group 1, whereas hatted quantities refer to gauge group 2.} $A_+=A+\hat A$, and are charged under the anti-diagonal gauge field $A_-=A-\hat A$. Also, it is important to note that this theory has no superpotential for the one D3-brane case we are considering.
In order to determine the moduli space, we need to solve the D-term equations (since there's no superpotential, there are no F-term equations to impose). The D-terms for this theory can be written as:
\bea
\frac{D^2}{g^2} + D \left(|z_1|^2+|z_2|^2-|w_1|^2-|w_2|^2\right),
\label{Dconi}
\eea
where $D$ is the usual auxiliary field. Integrating out the auxiliary field $D$, the D-terms can be written as (neglecting overall numerical factors):
\bea
g^2\left(|z_1|^2+|z_2|^2-|w_1|^2-|w_2|^2\right)^2.
\eea
The moduli space of vacua is obtained by imposing the vanishing of the D-terms and dividing by the gauge group. Thus, we need to impose the condition:
\bea
|z_1|^2+|z_2|^2-|w_1|^2-|w_2|^2 = 0,
\label{coniI}
\eea
and the identifications:
\bea
z_i \sim z_i e^{i \alpha},\;\;\;\;\;\;\;\;\;\;w_i \sim w_i e^{-i \alpha}.
\label{identi}
\eea
The moduli space of this gauge theory is precisely the conifold. This manifold can be viewed as a cone over $(SU(2)\times SU(2))/U(1)$, which in the literature is often referred to as $T^{1,1}$. The two $SU(2)$'s can be thought of as flavour symmetries, one acting on the $z_i$'s, which transform as a doublet, and the other acting on the $w_i$'s, transforming as a doublet. Also, the $U(1)$ group is the interacting anti-diagonal subgroup $A_-$ under which the chiral fields are charged.

Another way to define the conifold is by introducing four complex variables, $u_{ij}$, defined as:
\bea
&& u_{11} = u_1 + iu_2,\qquad u_{12} = iu_3 - u_4,\nn \\
&& u_{21} = iu_3 + u_4,\qquad u_{22} = u_1 - iu_2,
\eea
related to the chiral fields by the relation
$u_{ij} = z_i w_j$.
In terms of the new variables, the defining equation for the conifold can be written simply as:
\bea
\det u_{ij} = 0.
\label{coniII}
\eea
This way of defining the conifold is equivalent to the one we have discussed above. Given the definition of the coordinates $u_{ij}$ in terms of $w_i$ and $z_i$, the equation (\ref{coniII}) remains the same if we act on the chiral fields with the transformations:
\bea
z_i \rightarrow \lambda z_i \qquad w_i \rightarrow \lambda^{-1}w_i\qquad \lambda \in \BC^*.
\eea
If we write $\lambda = s e^{i\alpha}$, with $s \in \BR^+$ and $\alpha \in \BR$, the parameter $s$ can be used to satisfy (\ref{coniI}), while $\alpha$ parameterizes the gauge invariance (\ref{identi}).

A generalization to the case where we have a stack of $N$ parallel D3-branes probing the tip of the cone is straightforward. In that case, we have a Yang-Mills theory with $\mathcal{N}=1$ supersymmetry, gauge group $U(N)_1\times U(N)_2$, and chiral matter fields which transform under these groups according to Table \ref{t:chargeconiN}.\\
\begin{table}[h!]
 \begin{center}  
  \begin{tabular}{|c||c|c|}
  \hline
  \;& $U(N)_1$&$U(N)_2$\\
  \hline\hline  

  $z_1, z_2$&$  \textbf{N}$&$ \bar{\textbf{N}}$ \\
  \hline
  $w_1, w_2$&$  \bar{\textbf{N}}$&$ \textbf{N}$ \\
  \hline
\end{tabular}
  \end{center}
\caption{Transformations of the chiral fields under the two gauge groups. The symbol $\textbf{N}$ indicates that a field transforms in the fundamental representation of $U(N)$, whereas $\bar{\textbf{N}}$ indicates that a field transforms in the anti-fundamental representation of $U(N)$.}
\label{t:chargeconiN}
\end{table}

\begin{figure}[htbp]
\begin{center}
\includegraphics[scale=0.61]{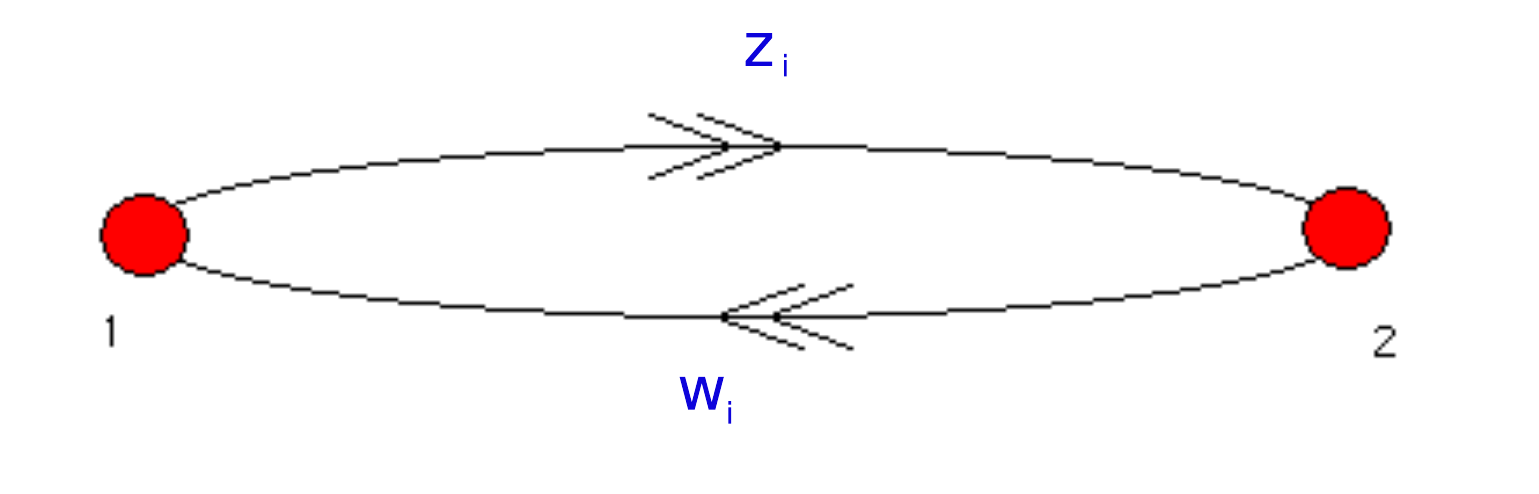}
\caption{The quiver diagram for the conifold gauge theory.}
\label{fig:conifold}
\end{center}
\end{figure}

The matter content of this theory can be represented with a so called ``quiver'' diagram, which we show in Figure \ref{fig:conifold}.
The red nodes correspond to the gauge groups, and each arrow corresponds to a chiral superfield of the theory. An outgoing (incoming) arrow from a given node signifies that the corresponding chiral field transforms in the fundamental (anti-fundamental) representation of the corresponding gauge group. Arrows starting and ending on the same node transform in the adjoint representation of the corresponding gauge group.

For the case with multiple branes, a superpotential needs to be added. The only superpotential quartic in the superfields
and invariant under the $SU(2)\times SU(2)$ flavor symmetry is:
\bea
W \propto \tr\left(\epsilon^{AC}\epsilon^{BD} z_A w_B z_C w_D\right) \propto \tr\left(z_1 w_1 z_2 w_2 - z_1 w_2 z_2 w_1\right)\ .
\label{supconi}
\eea
Note that this superpotential vanishes in the abelian case, where all the fields commute.
All the fields have R-charge equal to $1/2$. Thus, at the conformal fixed point, the scaling dimension of each field is:
\bea
\Delta = \frac{3}{2} |R| = \frac{3}{4}.
\eea
It can be shown that the superpotential is an exactly marginal operator for this gauge theory.

The gauge group of the theory, $U(N)_1\times U(N)_2$, contains two $U(1)$ factors. The diagonal subgroup of the gauge group, under which all the fields are neutral, decouples trivially.
The anti-diagonal subroup, commonly denoted as $U(1)_B$, becomes a global symmetry far in the IR because its gauge coupling flows to zero. Therefore,
the AdS/CFT correspondence for this case states that Type IIB string theory on $AdS_5 \times T^{1,1}$ with $N$ units of RR flux on $T^{1,1}$ is dual to $\mathcal{N}=1$ supersymmetric Yang-Mills theory with gauge group $SU(N)_1 \times SU(N)_2$.

\section{One M2-brane and Abelian Chern-Simons Theory}
Having briefly introduced the AdS/CFT correspondence in the context of D3-branes, we are now ready to discuss the correspondence for M2-branes.
Let us recall the formulation of 3-dimensional $\mathcal{N}=2$ supersymmetric Chern-Simons theories coupled to charged fields \cite{Schwarz:2004yj,Gaiotto:2007qi}.
In the superspace formalism, the Chern-Simons action for theories with a single $U(1)$ gauge group and $N_f$ matter flavours can be written, in Wess-Zumino gauge, as:
\bea
S = \int \ud^3 x \int \ud^4\theta \left(\frac{k}{4\pi} \sV \Sigma + \sum^{N_f}_{i=1}\Phi_i e^{q_i \sV} \Phi_i\right),\qquad k\in \BZ
\eea
where $\Phi_i$ are chiral matter superfields transforming under the gauge group with charge $q_i$, and where:
\bea
\sV &=& 2i \theta \bar{\theta} \sigma  + 2 \theta \gamma^\mu \bar{\theta} A_{\mu} + \sqrt{2}i \theta^2\bar{\theta} \bar{\chi} - \sqrt{2}i \bar{\theta}^2 \theta \chi + \theta^2 \bar{\theta}^2 D,\nn \\
\Sigma &=& \bar{D}^{\alpha}D_{\alpha} \sV.
\eea
As we can see from above, the vector superfield $\sV$ is composed of a gauge field $A_\mu$, a two-component Dirac spinor $\chi$, a scalar field $\sigma$, which comes from the $A_3$ component of the gauge field when we do dimensional reduction from the 3+1 dimensional theory, and another scalar field $D$.

The parameter $k$ is called the Chern-Simons level: the requirement that a non-abelian theory is invariant under large gauge transformations restricts it to integer values.
In components, the action can be written as:
\bea
S_{CS} &=& \frac{k}{4 \pi} \int \ud^3 x \;\left(\epsilon^{\mu\nu\lambda}A_{\mu}\partial_{\nu}A_{\lambda} + i \bar{\chi}\chi - 2 D \sigma\right),\nn \\
S_{matter} &=& \int \ud^3 x \; \sum^{N_f}_{i=1} \Bigl(-\sD_{\mu} \phi^{\dagger}_{i} \sD^{\mu} \phi_i - i \zeta^{\dagger}_i \slashed{\sD} \zeta_i + q_i \phi^{\dagger}_i D \phi_{i} - q^2_i \phi^{\dagger}_i \sigma^2 \phi_i +\nn \\
&& - q_i\zeta^{\dagger}_{i}\sigma \zeta_i + i q_i\phi^{\dagger}_i\bar{\chi} \zeta_i - i q_i \zeta^{\dagger}_i \chi \phi_i \Bigr),
\eea
where $\sD_\mu$ represents the covariant derivative, and where $\phi_i$ and $\zeta_i$ represent, respectively, the scalar and the fermionic part of the chiral matter field $\Phi_i$.\\
Note that all the fields in the vector multiplet are non-dynamical, so they are all auxiliary fields. Integrating out the scalar field $D$, we have that:
\bea
\sigma = -\frac{2\pi}{k} \sum^{N_f}_{i=1} q_i \phi^{\dagger}_i \phi_i,
\label{sextic}
\eea
and the D-term potential can be written as:
\bea
V_D \propto \sum^{N_f}_{i=1} q^2_i \phi^{\dagger}_i \phi_i \sigma^2.
\eea
Note that, because of (\ref{sextic}), the D-term potential is sextic in the scalar fields $\phi_i$.

Now let us consider the specific gauge theory proposed in \cite{Aharony:2008ug} as a description of a single M2-brane. This theory has gauge group $U(1)\times U(1)$, Chern-Simons levels $(k,-k)$ and four chiral superfields transforming under these groups as given in Table \ref{t:chargeconi}. Interestingly, the quiver diagram for this Chern-Simons gauge theory looks exactly like that in Figure \ref{fig:conifold}.
The matter action for this theory can be written in the superspace formalism as:
\bea
S_{matter} = \int \ud^3 x \int \ud^4 \theta \;\;\left(\bar{\mathcal{Z}}_A e^{-\sV}\mathcal{Z}^A e^{\hat{\sV}} + \bar{\mathcal{W}}^B e^{- \hat{\sV}} \mathcal{W}_B e^{\sV}\right),\qquad A,B=1,2,
\eea
where $\mathcal{Z}^A$ and $\mathcal{W}_B$ are chiral multiplets, whose lowest components are scalar fields that we denote as $Z^A$ and $W_B$.
Expanding the action in components, and deriving the equations of motion for the auxiliary fields, we have that:
\be
\sigma = \hat \sigma= \frac{2\pi}{k} \left(|Z_1|^2+|Z_2|^2-|W_1|^2-|W_2|^2\right)\ .
\label{sigmaabjm}
\ee
The D-term potential for this theory is proportional to $ \left(\sigma - \hat{\sigma}\right)^2 $; therefore, it vanishes.
Since the F-term potential vanishes as well (the abelian theory has no superpotential), we could be tempted to conclude that the moduli space is simply $\BR^8$, or $\BC^4$. However, this is true only up to a $\Integers_k$ identification.

Let us combine the fields as follows \cite{Aharony:2008ug,Benna:2008zy}:
\bea
Y^A = \left\{Z^A, W^{\dagger A}\right\}\qquad Y_A^{\dagger} = \left\{Z_A^{\dagger}, W_A\right\}.
\label{defya}
\eea
The newly defined fields $Y^A$ have the same charges as $Z^A$ under the gauge group $U(1)_1\times U(1)_2$.\\
With this definition, the bosonic part of the Chern-Simons action can be written as:
\bea
S_{\mathrm{bos}} = \frac{k}{4\pi}\int \ud^3 x\;\; \epsilon^{\mu\nu\lambda} \left(A_{\mu}\partial_{\nu}A_{\lambda} - \hat{A}_{\mu}\partial_{\nu}\hat{A}_{\lambda}\right) + \int \ud^3 x\;\; \sD_{\mu} Y^{\dagger}_{A} \sD^{\mu} Y^{A} \ .
\eea
Defining $A^{\pm}_{\mu} = A_{\mu} \pm \hat{A}_{\mu}$, the covariant derivative can be written as:
\bea
\sD_{\mu} Y^A = \partial_{\mu} Y^{A} + i (A_{\mu} - \hat{A}_{\mu}) Y^{A}=\partial_{\mu} Y^{A} + iA^{-}_{\mu} Y^{A}.
\eea
The CS action itself can be written in a more concise form as:
\bea
S_{\mathrm{CS}} = \frac{k}{4\pi}\int \ud^3 x\;\; \epsilon^{\mu\nu\lambda} A^{-}_{\mu} F^{+}_{\nu\lambda}.
\eea
If we define the theory on $\BR^3$ and map it to $\BR\times S^2$, then there are sectors with quantized monopole fluxes:\footnote{For a discussion of the extra factor of 2 appearing in the quantization condition, see \cite{Imamura:2008nn,Martelli:2008si}.}
\bea
\int_{S^2} F^{+} = 4\pi n\ .
\eea
If we now consider a gauge transformation for which:
\bea
A^{-}_{\mu} \rightarrow A^-_{\mu} + \partial_{\mu} \Lambda^-,\qquad Y^A\rightarrow e^{i\Lambda^-} Y^A,
\eea
this transformation brings about a boundary term that can be written as:
\bea
\delta S = \frac{k}{4	\pi} \Lambda^- \int_{S^2} F^+.
\eea
Since the fluxes are quantized, in order for the action to shift by $2\pi$ times an integer, we must require that:
\bea
\label{breaking}
\Lambda^- = \frac{2\pi l}{k} \ , \qquad l \in \Integers \ .
\eea
This provides an identification on the matter fields which implies that the moduli space for this theory with Chern-Simons levels $(k,-k)$ is $\BC^4/\BZ_k$.

Another way to derive this result is to note that, since $A^+_{\mu}$ does not appear in the action, we can treat $F^+_{\mu\nu}$ as a basic variable, rather than $A^+_{\mu}$ itself. Thus, in order to ensure that the equation $\ud F^{+}=0$ remains valid, we need to add a Lagrange multiplier \cite{Distler:2008mk,Lambert:2008et}:
\bea
S_{\tau} = \frac{1}{4\pi} \int \ud^3 x\;\; \tau \epsilon^{\mu\nu\rho} \partial_{\mu} F^+_{\nu\rho}.
\eea
Having inserted this term into the action, the equations of motion of $F^+_{\nu \rho}$ can be written as:
\bea
A^{-}_{\mu} = \frac{1}{k} \partial_{\mu} \tau \ .
\eea
The quantization condition on the fluxes requires that $\tau $ be periodic with period $2\pi$.
It follows that, under the gauge transformations,
\be
\tau \rightarrow \tau + k\Lambda^- \ .
\ee
We can use this gauge transformation to fix $\tau$ to be 0 but, because of its periodicity, we still have the freedom to make a transformation with $\Lambda^-=\frac{2\pi}{k}$. Examining how these ``large'' $\Integers_k$ gauge transformations act on the fields $Y^A$, we  again conclude that the moduli space of this Chern-Simons theory is $\BC^4/\BZ_k$.

In addition to the $\Integers_k$ gauge symmetry, the model has a global $U(1)_b$ symmetry which corresponds to the conserved current
$j_\mu \sim \epsilon_{\mu \nu \lambda} F^{+\nu\lambda}$. This symmetry acts by shifting the dual scalar $\tau$.
The monopole (anti-monopole) operators $e^{\pm i \tau}$, that are charged under the $U(1)_b$, create field configurations with magnetic flux of $F^+$ through the $S^2$ surrounding the point of insertion.

The non-abelian, $U(N)\times U(N)$ version of the model we are discussing was proposed in 2008 by Aharony, Bergman, Jafferis and Maldacena in \cite{Aharony:2008ug}, and in the literature is often referred to as the ABJM theory. It has the merit of having the conformal symmetry manifest, although the complete amount of supersymmetry is not manifest for $k=1,2$ (we will return to this issue in section 5).
A different, older description of the gauge theory on a stack of $N$ M2-branes is in terms of the IR limit of the gauge theory on $N$ D2-branes, i.e. the ${\mathcal N}=8$ supersymmetric Yang-Mills theory in $2+1$ dimensions.
For example, for a single D2-brane the action is:
\bea
S = \int \ud^3 x\;\left(-\frac{1}{4} F^2_{\mu\nu} - \sum^{7}_{i=1} \frac{1}{2} (\partial_{\mu}\phi^i)^2\right).
\eea
Dualizing the field strength to a scalar,
\bea
F_{\mu\nu}\sim \epsilon_{\mu\nu\lambda}\partial^{\lambda}\phi^8,
\eea
the $SO(8)$ symmetry acting on the 8 scalars is manifest. However, it is not known how to generalize the duality transformation to the non-abelian gauge theory on multiple D2-branes. In this approach to M2-branes, the ${\mathcal N}=8$ supersymmetry is manifest, but the  conformal invariance is not; it contains the coupling constant $g_{YM}$ which has dimension of Energy$^{1/2}$.
Presumably, the superconformal M2-brane theory is the IR sector of the theory on $N$ D2-branes which emerges for energies much smaller than $g^2_{YM}$.

\section{Non-Abelian Chern-Simons theory}

 A breakthrough in the search for a Chern-Simons matter theory with $\mathcal{N}=8$ supersymmetry came with the work of Bagger and Lambert \cite{Bagger:2006sk} and, independently, Gustavsson\cite{Gustavsson:2007vu}. They constructed a theory, often referred to as the BLG model, using a so-called ``3-algebra.'' Given a set of generators $T_a$, such an algebra can be defined by introducing the triple product:
\bea
\left[T^a,T^b,T^c\right] = f^{abc}_{\ \ \ d} T^d,
\eea
where $f_{abcd}$ is a fully anti-symmetric tensor. Given this algebra, a maximally supersymmetric Chern-Simons lagrangian is:
\bea
\mathcal{L}_{CS} &=& \frac{1}{2} \epsilon^{\mu\nu\lambda}\left(f^{abcd}A_{\mu a b}\partial_{\nu}A_{\lambda c d} + \frac{2}{3} f^{cda}_{\ \ \ g}f^{efgb} A_{\mu a b}A_{\nu c d}A_{\lambda e f}\right),\nn \\
\mathcal{L}_{matter} &=& -\frac{1}{2} \sD^{\mu} x^{a I} \sD_{\mu} x^{I}_{a} + \frac{i}{2} \bar{\psi}^a \Gamma^{\mu} \sD_{\mu} \psi_a + \frac{i}{4}\bar{\psi}_b \Gamma_{IJ} x^{I}_{c} x^{J}_{d} \psi_a f^{abcd} \nn \\
&&- \frac{1}{12} \tr \left([x^I,x^J,x^K][x^I,x^J,x^K]\right),\qquad I,J=1,\ldots,8,
\eea
where $A^{\mu}_{ab}$ is the gauge boson, and $\psi_a$ and $x^I = x^I_a T^a$ are matter fields\footnote{By convention, space-time indices are denoted by Greek letters $\mu,\nu,\lambda,\ldots$, gauge indices by $a,b,c,\ldots$, and SO(8) vector indices by capital letters $I,J,\ldots$.}. If we let $a=1,\ldots,4$, then we can obtain an $SO(4)$ gauge symmetry by choosing $f^{abcd}=f\epsilon^{abcd}$, $f$ being a constant.\footnote{The invariance under large gauge transformations requires $f = \frac{2\pi}{k}$.} Not only does this seem like a natural choice, but it turns out to be the only one that gives a gauge theory with manifest unitarity and $\mathcal{N}=8$ supersymmetry.

After this model was proposed, it was shown \cite{Bandres:2008vf, VanRaamsdonk:2008ft} that this gauge theory is equivalent to an $SU(2)\times SU(2)$ Chern-Simons gauge theory with opposite Chern-Simons levels:
\bea
\mathcal{L}_{CS} &=& \frac{k}{4\pi} \epsilon^{\mu\nu\lambda} \tr\left(A_{\mu}\partial_{\nu}A_{\lambda} + \frac{2i}{3} A_{\mu}A_{\nu}A_{\lambda} - \hat{A}_{\mu}\partial_{\nu}\hat{A}_{\lambda} - \frac{2i}{3} \hat{A}_{\mu}\hat{A}_{\nu}\hat{A}_{\lambda}\right),\nn \\
\mathcal{L}_{matter} &=& -(\sD^{\mu}X^I)^{\dagger} \sD_{\mu}X^I + i\bar{\Psi}^{\dagger} \Gamma^{\mu}\sD_{\mu}\Psi - \frac{4i\pi}{k}\bar{\Psi}^{\dagger}\Gamma^{IJ}\Big(X^I X^{J\dagger}\Psi + X^J\Psi^{\dagger}X^I+\nn \\
&&+ \Psi X^{I\dagger}X^J\Big) - \frac{32\pi^2}{3k^2}\tr \left(X^{[I}X^{\dagger J}X^{K]}X^{\dagger [K}X^{J}X^{\dagger I ]}\right),
\label{vanraam}
\eea
where the covariant derivative can be written as:
\bea
\sD_{\mu} X^I = \partial_{\mu} X^I + iA_{\mu} X^I - i X^I \hat{A}_{\mu}.
\eea
The bi-fundamental matter fields $X^I$ are related to the ones in the $SO(4)$ notation by:
\bea
X^I = \frac{1}{2}(x^I_4 \BI_{2\times 2} + i x^I_i \sigma^i),
\label{relso4su2su2}
\eea
where $\BI_{2\times 2}$ is the $2\times 2$ identity matrix and $\sigma^i$ are the Pauli matrices.\\
Also, the bi-fundamental matter fields satisfy the reality condition \cite{VanRaamsdonk:2008ft}:
\bea
X^{\hat{a}}_{a} = -\epsilon_{ab}\;X^{b}_{\hat{b}}\;\epsilon^{\hat{a}\hat{b}}, \qquad \epsilon^{ab} = i\sigma^{ab}_2.
\label{realcond}
\eea

  An $\mathcal{N}=2$ superspace formulation of this theory was given in \cite{Aharony:2008ug,Benna:2008zy}. The gauge fields $A$ and $\hat{A}$ can be thought of as components of two vector supermultiplets, $\mathcal{V}$ and $\hat{\mathcal{V}}$, whose other components are auxiliary fields.
The matter fields can be combined into chiral bi-fundamental superfields, which we denote as $\mathcal{Z}^A$. The lowest component of $\mathcal{Z}^A$ is a complex scalar field $Z^A$, related to the $X^I$ fields by:
\bea
Z^A = X^A + iX^{A+4},\qquad A=1,\ldots,4.
\label{relvanraam}
\eea
In terms of  the chiral multiplets $\mathcal{Z}^A_a$ transforming under the $SO(4)$ gauge group and carrying R-charge $1/2$, the superpotential can be written as:
\bea
W = -\frac{\pi}{4! k} \epsilon_{ABCD}\epsilon^{abcd}(\mathcal{Z}^A_a \mathcal{Z}^B_b \mathcal{Z}^C_c \mathcal{Z}^D_d),
\label{supso4}
\eea
  Thus, the theory has a manifest $U(1)_R \times SU(4)$ symmetry. By virtue of (\ref{relso4su2su2}), this superpotential can be expressed in terms of the bi-fundamental chiral superfields $\mathcal{Z}^A$ as:
\bea
W = \frac{\pi}{3 k} \epsilon_{ABCD} (\mathcal{Z}^A \mathcal{Z}^{\ddagger B} \mathcal{Z}^C \mathcal{Z}^{\ddagger D})\ ,
\label{supsu2su2}
\eea
where we have introduced the operations:
\bea
Z^{\ddagger A} &=& -\epsilon (Z^A)^{\trans} \epsilon = X^{\dagger A} + i X^{\dagger A+4},\nn \\
\bar{Z}_{A} &=& -\epsilon (Z^A)^{*} \epsilon = X^{A} - i X^{A+4}.
\label{ddag}
\eea
Although the first of these operations might look like a hermitian conjugation, we note that there is no conjugation of the imaginary part, but only of the scalar multiplets $X^I$. This operation is crucial because it does not break the holomorphy of the superpotential.

It is not obvious that this $\mathcal{N}=2$ formalism describes the BLG model. The latter has $\mathcal{N}=8$ supersymmetry, and therefore an $SO(8)_R$ invariance, whereas (\ref{supsu2su2}) only has a manifest $U(1)_R\times SU(4)$ symmetry. However, expressing the action of this model in components and integrating out all the auxiliary fields, the sum of the D-term potential and of the F-term potential has the desired $SO(8)_R$ global symmetry \cite{Benna:2008zy}.

Since the reality condition (\ref{realcond}) and the ``double dagger'' operation (\ref{ddag}) are special to $SU(2)\times SU(2)$, it seems difficult to generalize the construction to gauge groups with a higher rank.
A way to overcome this difficulty, proposed in \cite{Aharony:2008ug}, is to abandon the manifest global $SU(4)$ invariance by forming the following combinations of the bi-fundamental fields:
\bea
Z^1 &=& X^1 + iX^5, \qquad W^1 = X^{3\dagger} + iX^{7\dagger}\ ,\nn \\
Z^2 &=& X^2 + iX^6, \qquad W^2 = X^{4\dagger} + iX^{8\dagger}\ .
\eea
If we promote the fields $Z^A$ and $W^A$ to chiral superfields, that we shall denote as $\mathcal{Z}^A$ and $\mathcal{W}^A$, the superpotential of the model can be written as \cite{Aharony:2008ug,Benna:2008zy}:
\be
W = \frac{2\pi}{k}\epsilon_{AC}\epsilon^{BD}\tr \left(\mathcal{Z}^A\mathcal{W}_B\mathcal{Z}^C\mathcal{W}_D\right)\ .
\ee
This superpotential looks exactly the same as that for the D3-brane theory on the conifold and can be easily generalized to higher rank gauge groups of the type $SU(N)\times SU(N)$. Furthermore, there is an important \textit{caveat}. Apart from the manifest $U(1)_R\times SU(2)\times SU(2)$ global symmetry, the superpotential has also a $U(1)$ symmetry, under which the chiral multiplets transform as:
\bea
\mathcal{Z}^A\rightarrow e^{i\alpha} \mathcal{Z}^A, \qquad \mathcal{W}_B \rightarrow e^{-i\alpha}\mathcal{W}_B.
\eea
In the $3+1$ dimensional theory on $N$ D3-branes at the conifold singularity, this starts out as a gauge symmetry but in the far IR becomes global and is identified with a $U(1)$ baryonic symmetry. In the $2+1$ dimensional case at hand, the dynamics is different
and we have to treat this $U(1)$ as a gauge symmetry, although eventually it becomes broken to a $\Integers_k$ subgroup (\ref{breaking}).
 Using this argument, as well as type IIB brane constructions, ABJM proposed that the gauge group on $N$ M2-branes is $U(N)\times U(N)$, and not $SU(N)\times SU(N)$. We have already checked that the moduli space of the $N=1$ abelian theory with Chern-Simons levels $(k,-k)$ is the $\BZ_k$ orbifold of $\BC^4$; thus, this theory correctly describes a single M2-brane. Additional geometrical arguments for why the ABJM theory describes $N$ M2-branes on $\BC^4/\BZ_k$ are presented in \cite{Aganagic:2009zk,Davey:2009qx}. A type IIA reduction performed there leads to D2-branes on the conifold fibered over $\BR$, with a 2-form RR-flux turned on.
 This explains the close relation between the ABJM theory and the theory \cite{Klebanov:1998hh} for D-branes on the conifold.

An important property of the classical ABJM action is that the $U(1)_R\times SU(2)\times SU(2)$ global symmetry
 is enhanced to $SU(4)_R$, corresponding to $\mathcal{N}=6$ supersymmetry.
The scalar potential is made of two parts, one coming from the F-terms and the other from the D-terms. The former can be written as:
\bea
V^{\mathrm{bos}}_F &=& \left|\frac{\partial W}{\partial Z^A}\right|^2 + \left|\frac{\partial W}{\partial W_A}\right|^2 = \tr \left[F^{\dagger}_A F^A + G^{\dagger A} G_A\right],\nn \\
F^A &=& \frac{4\pi}{k} \epsilon^{AC}\epsilon_{BD} W^{\dagger B}Z^{\dagger}_{C}W^{\dagger D},\nn \\
G_A &=& \frac{4\pi}{k} \epsilon_{AC}\epsilon^{BD} Z^{\dagger}_{B} W^{\dagger C} Z^{\dagger}_{D}.
\eea
The contribution coming from the D-terms can be written as:
\bea
V^{\mathrm{bos}}_D &=& \tr \left[N^{\dagger}_{A} N^{A} + M^{\dagger A}M_{A}\right], \nn \\
N^A &=& \sigma Z^A - Z^A \hat{\sigma}, \qquad \qquad \qquad
M_A = \hat{\sigma}W_A - W_A \sigma, \nn \\
\sigma &=& \frac{2\pi}{k} (Z^A Z_A^{\dagger} - W^{\dagger B} W_B),  \qquad
\hat{\sigma} = \frac{2\pi}{k} (Z_A^{\dagger}Z^A - W_B W^{\dagger B}).
\eea
Combining the two contributions and using the notation introduced in (\ref{defya}), we find that the full bosonic potential is:
\bea
V^{\mathrm{bos}} &=& -\frac{4\pi^2}{3 k^2} \tr \Bigsbrk {Y^A Y^\dagger_A Y^B Y^\dagger_B Y^C Y^\dagger_C + Y^\dagger_A Y^A Y^\dagger_B Y^B Y^\dagger_C Y^C +\nn \\
&& + 4 Y^A Y^\dagger_B Y^C Y^\dagger_A Y^B Y^\dagger_C  - 6 Y^A Y^\dagger_B Y^B Y^\dagger_A Y^C Y^\dagger_C }.
\eea
The interaction terms, quadratic in the fermion fields and quartic in the scalars, possess the manifest
$SU(4)_R \sim SO(6)_R$ symmetry as well \cite{Benna:2008zy}. Therefore, this symmetry is manifest in the classical action; it strongly suggests that, for general $N$ and $k$, the ABJM theory has at least $\mathcal{N}=6$ supersymmetry. An explicit demonstration of the
$\mathcal{N}=6$ superconformal invariance of the ABJM theory was presented in \cite{Bandres:2008ry}.

\section{Gravitational description of coincident M2-branes}

 A stack of $N$ coincident M2-branes creates the following extremal geometry:
\bea
d s^2_{11} &=& h(r)^{-2/3}\left(-dt^2 + dx^2_1 + dx^2_2\right) + h(r)^{1/3}\left(dr^2 + r^2d\Omega^2_7\right),\nn \\
h(r) &=& 1 + \frac{L^6}{r^6}\ ,\qquad L^6 = 32 \pi^2 N l^6_p,\nn \\
F_4 &=& d^3x \wedge dh(r)^{-1}.
\label{metricm2}
\eea
In the limit where $r\rightarrow 0$, the metric becomes that of $AdS_4 \times S^7$:
\bea
d s^2_{11} = L^2\left( \frac{1}{4} d s^2_{AdS_4} + d s^2_{S^7}\right).
\eea
For the strongly coupled theory on $N$ M2-branes, the AdS/CFT correspondence predicts an interesting essential feature that has not been completely understood yet: the number of degrees of freedom scales as $N^{3/2}$ \cite{Klebanov:1996un}, not as $N^2$ found for the strongly coupled gauge theories on D3-branes.
One way to see this is through studying the correlation functions of protected gauge invariant operators.
 Since their spectrum is in one-to-one correspondence with the Kaluza-Klein harmonics on $S^7$, we can use the gravity side of AdS/CFT to predict that all their correlation functions scale like:
\bea
\langle \mathcal{O}_{1}\mathcal{O}_{2}\ldots \mathcal{O}_{m}\rangle \propto (L/l_p)^9 \propto N^{3/2}\ ,
\eea
where we have used the relation between $L$ and $N$ obtained in (\ref{metricm2}).
Another way is to study the thermal gauge theory \cite{ Aharony:2008ug,Klebanov:2000me,Klebanov:1996un}. The non-extremal geometry for a stack of $N$ M2-branes is:
\bea
d s^2_{11} &=& h(r)^{-2/3} \left(-f(r) dt^2 + dx^2_1 + dx^2_2\right) + h^{1/3}(r)\left(\frac{dr^2}{f(r)} + r^2 d\Omega^2_7\right),\nn \\
h(r) &=& 1 + \frac{L^6}{r^6},\qquad
f(r) = 1 - \frac{r^6_0}{r^6} \ .
\eea
In the near-horizon region, $r\ll L$, this becomes a black brane in $AdS_4$, and
$r_0$ represents the Schwarzschild radius, which is related to the Hawking temperature $T$.

If we consider a stack of M2-branes placed at the singularity of the orbifold $\BC^4/\BZ_k$, which is described by the level $k$ ABJM theory, then $d\Omega^2_7$ refers to the metric on unit $S^7/\BZ_k$.
When the 't Hooft coupling $\lambda = N/k$ is very large, so that the metric is weakly curved, the Bekenstein-Hawking entropy is written as:
\bea
S_{BH} = \frac{A}{4G_{N}} = 2^{7/2}3^{-3}\pi^2 V_2 T^2 k^{1/2} N^{3/2} + O(N^{1/2}),
\eea
where $V_2$ is the spatial volume of the stack of M2-branes.
Thus, the AdS/CFT correspondence predicts that for large $\lambda$ the number of degrees of freedom in the ABJM theory scales like $N^{3/2}k^{1/2}$.
More generally, the thermal entropy of the ABJM theory can be written as:
\bea
\frac{S_{BH}}{V_2 T^2} \propto N^2 f(\lambda)\ .
\label{bhentro}
\eea
In the perturbative regime, $\lambda \ll 1$, we expect $ f(\lambda) = 1 + O(\lambda^2)$, while the AdS/CFT correspondence predicts that:
\bea
f(\lambda) \propto \frac{1}{\sqrt{\lambda}}, \qquad \lambda \gg 1.
\eea
This behavior is very different from the situation where a stack of D3-branes is considered. In that case the Bekenstein-Hawking entropy can be written in a similar form to (\ref{bhentro}) but, when the 't Hooft coupling becomes large, the function $f(\lambda)$
approaches $3/4$ at large $\lambda$.\footnote{The situation becomes even more interesting when we consider a stack of M5-branes, because the supergravity approximation shows that the Bekenstein-Hawking entropy scales like $N^3$ \cite{Klebanov:1996un}.}

Now, let us discuss the parity transformation. In $2+1$ dimensions, the reflection of both coordinates is simply a rotation; therefore, one has to consider the reflection of only one. Sometimes it has to be accompanied by the action on other fields.
For example, in $AdS_4 \times Y^7$, the parity transformation acts on one of the spatial coordinates and on the 3-form:
\bea
x^1\rightarrow -x^1,\qquad C_3\rightarrow -C_3\ .
\eea
 If only the $C_{012}$ component of the 3-form is turned on, this transformation preserves the background.
This shows that the dual gauge theory cannot be a supersymmetric Chern-Simons theory with a single gauge group $U(N)$, where the Chern-Simons term violates the parity \cite{Schwarz:2004yj}. However, theories with two gauge groups and opposite
Chern-Simons levels preserve a parity symmetry that involves the interchange of the gauge groups \cite{Bandres:2008vf,Deser:1981wh}.
For example, in the ABJM theory, the parity transformation
\bea
x^1 \rightarrow -x^1, &\qquad&
A\leftrightarrow \hat{A},\nn \\
Y^A \leftrightarrow Y^{\dagger}_A,&\qquad& \psi_\alpha^{\dagger A} \leftrightarrow \gamma^1_{\alpha\beta}\psi_A^\beta,
\eea
is a symmetry of the action. This is consistent with the parity symmetry of the $AdS_4\times S^7/\Integers_k$ background of M-theory.

\section{Supersymmetry Enhancement and Monopole Operators}

Let us show that $\mathcal{N}=6$ is the correct amount of supersymmetry
for $N$ M2-branes placed at the singularity of $\BC^4/\BZ_k$ for $k>2$. The $\BZ_k$ orbifold acts on the 4 complex coordinates of $\BC^4$ as:
\bea
y^A \rightarrow e^{2\pi i / k} y^A.
\eea
Note that this preserves the $SU(4)$ symmetry that rotates the $y^A$; this is the $R$-symmetry from the gauge theory point of view.
The generators of $\BZ_k$ act on the $SO(8)$ spinors as:
\bea
\Psi \rightarrow e^{2\pi i (s_1+s_2+s_3+s_4)/k}\Psi,
\eea
where $s_i = \pm 1/2$ are the spinor weights. By chirality projection, the sum of the $s_i$'s has to be even, giving an 8 dimensional representation. Therefore, the spinors that survive the orbifold projection must satisfy:
\bea
\sum^{4}_{i=1} s_i = 0 \;\;\;(\mathrm{mod}\;\; k).
\eea
 For $k>2$, 6 of the 8 spinors are left invariant by the orbifold action. It follows that this supersymmetric gauge theory is expected to have 12 supercharges and, accordingly, $\mathcal{N}=6$ supersymmetry, in agreement with that in the classical action of the
 ABJM theory. However, for $k=1,2$ all the spinors are found to be invariant, and the supersymmetry is enhanced to
$\mathcal{N}=8$.
In the following, we will investigate what causes this supersymmetry enhancement in the ABJM model.

 In the ABJM model, the classical global symmetry is $U(1)_b\times SU(4)_R$. The $SU(4)$ symmetry is realized by the 15 conserved traceless currents:
\bea \label{obvious}
j^{A}_{\mu B} = i\tr \left[Y^A \sD_{\mu} Y^\dagger_B - (\sD_{\mu} Y^A) Y^\dagger_B + i \psi^{\dagger A} \gamma_{\mu} \psi_B\right].
\eea
The $U(1)_b$ transformation acts as:
\be
Y^A \rightarrow e^{i\alpha} Y^A\ , \qquad \psi_B \rightarrow e^{i\alpha}\psi_B
\ .
\ee
The corresponding current is related by the $A^-$ equation of motion (in the $U(N)\times U(N)$ theory,
$A^\pm \sim \tr A \pm \tr \hat A$) to the current
$
j_\mu \sim \epsilon_{\mu \nu \lambda} F^{+\nu\lambda}\ ,
$
which is obviously conserved. Therefore, the $U(1)_b$ charge is carried only by the field configurations that have a flux of
$F^+$ through the 2-sphere at infinity of $\Reals^3$. Such field configurations are created by the so-called ``monopole operators'' \cite{Borokhov:2002ib} that we discuss next. For some of their recent applications see \cite{Aharony:2008ug,Gaiotto:2008ak,Berenstein:2008dc,Klebanov:2008vq,Imamura:2009ur,Gaiotto:2009tk,Kim:2009wb, Benna:2009xd, Gustavsson:2009pm}.

In order for the supersymmetry to be enhanced to $\mathcal{N}=8$, we need the global symmetry to be enhanced to $SO(8)_R$, which has 28 generators. Therefore, we need to find 12 conserved currents in addition to the 16 $U(1)_b\times SU(4)_R$ currents.
 Construction of these 12 currents is expected to involve the monopole operators \cite{Aharony:2008ug}. Each of these operators creates a quantized flux in a $U(1)$ subgroup of the gauge group through a sphere surrounding the insertion point.
In the ABJM theory, which has gauge group $U(N)_1\times U(N)_2$, these monopole operators are labeled by the Cartan generators $H$ and $\hat{H}$ of each of the two $U(N)$ factors of the gauge group:
\bea
H = \diag (q_1, q_2, \ldots, q_N),\qquad \hat{H} = \diag (\hat{q}_1, \hat{q}_2, \ldots, \hat{q}_N), \qquad q_i, \hat{q}_i \in \BN,
\eea
where the condition that the entries of the generators have to be integers follows from the flux quantization condition around the $S^2$ surrounding the insertion point of the monopole. Also, for convenience, one can arrange the $q_i$'s and the $\hat{q}_i$'s such that:
\bea
q_1 \geq q_2 \geq \ldots \geq q_N, \qquad  \hat{q}_1 \geq \hat{q}_2 \geq \ldots \geq \hat{q}_N.
\eea
We will restrict our attention to monopoles with $H=\hat H$. We also note that, for each monopole, there is an anti-monopole.

Generically, if a gauge group has Chern-Simons level $k$, the monopole operators transform in the representation of the gauge group given by a Young tableaux with $k q_1$ boxes in the first row, $k q_2$ boxes in the second row, etc\footnote{Since the gauge groups we are dealing with are $U(N)$ rather than $SU(N)$, the columns of length $N$ must be taken into account as well.}. It follows that in the ABJM theory, the monopoles are not gauge singlets and can combine with the chiral matter to form gauge invariant operators.

Consider, for example, the $k=1$ theory. Here, the simplest, unit-charge monopole operator corresponds to
$q_1 = \hat q_1 =1$. The corresponding operator, ${\mathcal M}^a_{\hat a}$, transforms as a fundamental under $U(N)_1$ and anti-fundamental under $U(N)_2$. There is also the anti-monopole operator that transforms in the conjugate representation,
$({\mathcal M}^{-1})^{\hat a}_a$. Similarly, there are doubly charged monopole operators corresponding for example to
 $q_1 = \hat q_1 =2$, that we denote $({\mathcal M}^2)^{ab}_{\hat a \hat b} $, which transform in the symmetric tensor representation under $U(N)_1$, and in the conjugate representation under $U(N)_2$. We will denote their anti-monopole operators by
 $({\mathcal M}^{-2})^{\hat a \hat b}_{a b} $. Another type of doubly charged BPS monopole operator has $q_1=q_2=\hat q_1= \hat q_2=1$. Such an operator transforms as an anti-symmetric tensor under $U(N)_1$, and in the conjugate representation under $U(N)_2$. As we increase the monopole charge, the variety of different monopoles increases.

Now, let us consider the current operators:
\bea
j^{AB}_{\mu} = i\left [ Y^A \sD_{\mu} Y^B - \sD_{\mu} Y^A Y^B + i \psi^{\dagger A} \gamma_{\mu} \psi^{\dagger B}\right ].
\eea
These operators are not gauge invariant, but they can be combined with the monopole $({\mathcal M}^{-2})^{\hat a \hat b}_{a b} $, which has $k q_1=2, k\hat{q}_1=2$ and $q_i=\hat{q}_i=0$ for $i\neq 1$, to form 6 invariant currents\footnote{The number 6 comes from the anti-symmetry of $j^{AB}_{\mu}$ under the exchange of $A$ and $B$.} that we shall denote as $J^{AB}_{\mu}$. Combining their complex conjugates $j_{AB\mu}$ with the monopole $\mathcal{M}^{2}$, gives another six conserved currents which are also gauge invariant. To summarize, the monopoles have provided us with 12 symmetry generators that we can add to the 16 obvious currents of $U(1)_b\times SU(4)_R$. With a total of 28 generators, the global symmetry is enhanced to $SO(8)_R$, as we expected.
When $k\geq 3$, there is no way to construct the monopole operators $\mathcal{M}^{-2}$ and $\mathcal{M}^{2}$, which carry two indices under each $U(N)$; so the symmetry enhancement does not happen.

The importance of the monopoles goes beyond the supersymmetry enhancement for $k=1,2$. They are also necessary for matching the spectrum of the gauge theory with that of the dual gravity theory.
In the ABJM theory with $k=1$, the simplest gauge invariant operators have the form $Y^{\dagger \hat a}_{B a} {\mathcal M}^a_{\hat a}$
and their conjugate. These operators, linear in the scalar fields, are expected to have scaling dimension $1/2$; they can be thought of as free fields dual to `singleton' modes in $AdS_4$.
The simplest scalar composite operators we can construct in the ABJM theory at level $k=1$ can be written as:
\bea
\tr\left[Y^{\dagger}_A Y^B - \frac{1}{4}\delta^B_A Y^{\dagger}_C Y^C\right]\ .
\eea
There are 15 such traceless operators of dimension 1. The AdS/CFT correspondence implies that the gauge invariant scalar operators should be in one-to-one correspondence with the Kaluza-Klein harmonics on $S^7$. There are 35 such harmonics that correspond to operators of dimension 1. Therefore, we need to find 20 more operators made of two scalar fields in order to match this part of spectrum with the gravity side. Once again, we proceed by forming non-gauge-invariant combinations of scalar fields and combining them with monopoles in order to make them gauge invariant. The 20 gauge invariant operators written as:
\bea
\label{composite}
Y^\dagger_A Y^\dagger_B \mathcal{M}^2, \qquad Y^A Y^B \mathcal{M}^{-2},
\eea
turn out to be what is needed to match the spectrum of the field theory with the gravity theory. These additional operators have non-vanishing $U(1)_b$ charge, which is dual to the momentum along the M-theory circle.\footnote{If we view the $S^7$ as a circle fibration over $\mathbb{CP}^3$, then the reduction to type IIA string theory produces an $AdS_4\times \mathbb{CP}^3$ background.}

\begin{figure}[htbp]
\begin{center}
\includegraphics[scale=0.50]{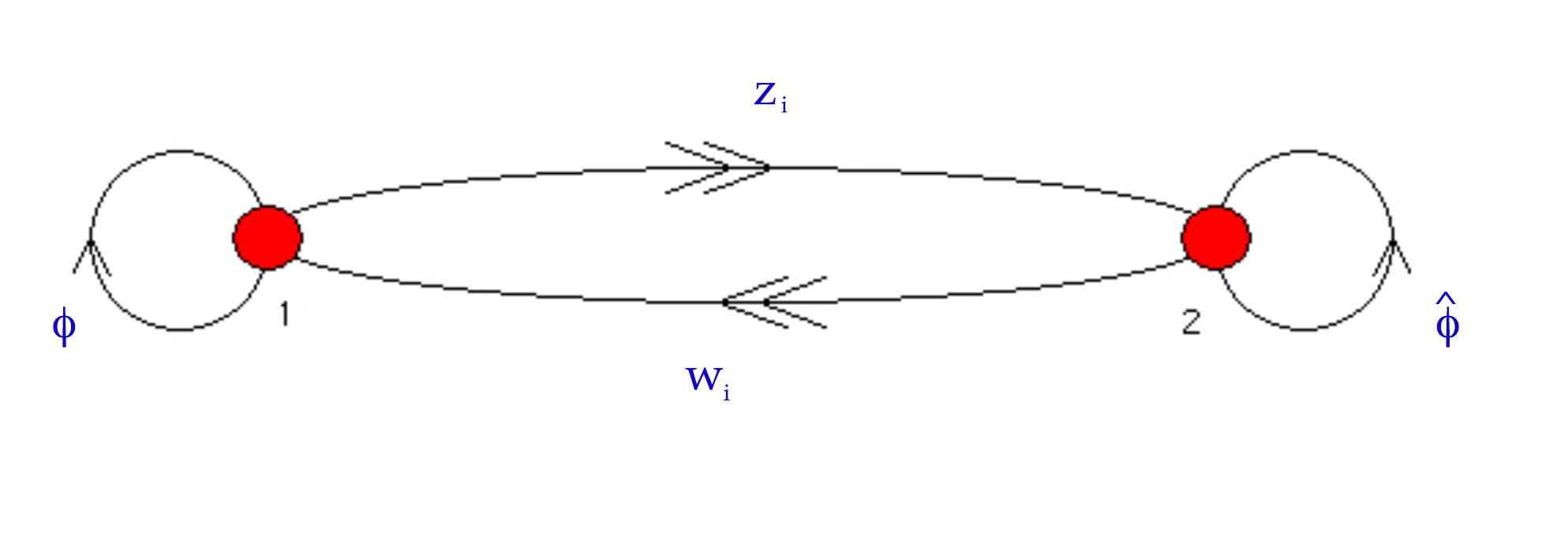}
\caption{The quiver diagram for the $\mathcal{N}=3$ supersymmetric gauge theory which flows to ABJM theory in the IR.}
\label{fig:conifoldadj}
\end{center}
\end{figure}

It remains to show that the monopole operators do not alter the `naive dimension' $1$ of the scalar bilinears.
Since the ABJM theory at low $k$ is a strongly coupled gauge theory, the monopole operators are quite hard to analyse in detail as there's no perturbative approach available. However, a way to overcome this problem is to embed the ABJM theory in an $\mathcal{N}=3$ supersymmetric Yang-Mills-Chern-Simons theory, by adding a Yang-Mills term for the gauge fields in the action \cite{Benna:2009xd}:
\bea
S_{YM} &=& \frac{1}{4g^2}\int \ud^3 x \int \ud^2 \theta\;\; \tr \left[\mathcal{U}^\alpha \mathcal{U}_\alpha+ \hat{\mathcal{U}}^\alpha \hat{\mathcal{U}}_\alpha\right],\nn \\
\mathcal{U}_\alpha &=& \frac{1}{4}\bar{D}^2 e^{\mathcal{V}} D_\alpha e^{-\mathcal{V}},
\eea
where $\mathcal{V}$ is the vector superfield. Also, we add two dynamical adjoint superfields with kinetic term:
\bea
S_{\mathrm{adj}} &=& \frac{1}{g^2} \int \ud^3 x \int \ud^2 \theta \tr\left[ -\bar{\Phi}e^{-\mathcal{V}}\Phi e^{\mathcal{V}} - \hat{\bar{\Phi}}e^{-\mathcal{\hat{V}}}\hat{\Phi} e^{\mathcal{\hat{V}}}\right]\ .
\label{kint}
\eea
The superpotential can be written as:
\bea
W = \tr\left(\Phi \mathcal{Z}^A\mathcal{W}_A + \hat{\Phi}\mathcal{W}_A\mathcal{Z}^A\right) + \frac{k}{8\pi} \tr\left(\Phi\Phi - \hat{\Phi}\hat{\Phi}\right).
\eea
The quiver diagram of this supersymmetric gauge theory is presented in Figure \ref{fig:conifoldadj}.\\
In the IR the gauge coupling $g$ flows to infinity, the kinetic term (\ref{kint}) vanishes and the adjoint fields can be integrated out. The resulting theory is precisely the ABJM model. In contrast, in the UV the gauge coupling goes to 0, and the theory becomes a weakly coupled $\mathcal{N}=3$ supersymmetric Yang-Mills theory that can be studied perturbatively. Therefore, we can compute quantities in the UV, where the theory is weakly coupled, and then study the flow to the IR, where the theory becomes the ABJM Chern-Simons theory. In particular, we can compute the non-abelian $SU(2)_R$ charge of the monopole operators, which is not modified by the RG flow \cite{Benna:2009xd}.

In the UV we may study the monopoles semi-classically.
In order for a classical monopole background to be BPS (or anti-BPS), we need to turn on the vacuum expectation values of the scalar fields $\phi_i$ that are in the vector multiplets and that transform in the spin 1 representation of the $R$-symmetry group:
\bea
\phi_i = -\hat{\phi} = \pm \frac{H}{2} n_i (\tau) \ .
\eea
The minus sign corresponds to a BPS background, while the plus to anti-BPS. The $n_i(\tau)$ vector is a unit vector on the $SU(2)_R/U(1)_R$ 2-sphere and it may vary with the Euclidean time $\tau$ adiabatically.

The vector $n_i(\tau)$ couples to the fermions; integrating them out we obtain the induced action for $n_i$. By solving for its quantum mechanical motion, we can determine the representations in which the monopoles transform under the $SU(2)_R$ symmetry. For the ABJM theory, the singlet representation turns out to be allowed.
Such monopole operators do not contribute to the R-charge and, therefore, to the scaling dimension of composite operators.
Thus, the dimensions of composite operators like (\ref{composite}) agree with the AdS/CFT correspondence.

\section*{Acknowledgments}
 We thank the organizers of the Galileo Galilei Institute summer school ``New Perspectives in String Theory'' for giving us the opportunity to participate in the school. I.R.K. is grateful to M. Benna, T. Klose, A. Murugan, M. Smedb\" ack, A. Tseytlin and E. Witten for collaboration on some of the material reviewed in these notes. G.T. is grateful to A. Hanany, N. Mekareeya and J. Davey for enlightening discussions on the topics discussed in these notes. We also thank M. Kiermaier and T. Klose for their useful comments on the manuscript. The work of I.R.K. was supported in part by the NSF grants PHY-0756966 and PHY-0551164. I.R.K. is grateful to the Kavli Institute for Theoretical Physics for hospitality during some of his work on these notes. G.T. is grateful to Giuni and Elisa Rebessi for their invaluable support during the preparation of this manuscript.

\bigskip

\end{document}